# Percolative Ferromagnetism in Anatase Co:TiO$_2$

S. R. Shinde[1,*], S. B. Ogale[1,2,¶], Abhijit S. Ogale[3], S. J. Welz[4], A. Lussier[5], Darshan C. Kundaliya[1], H. Zheng[2,#], S. Dhar[1], M.S.R. Rao[1], R. Ramesh[2,#], Y. U. Idzerda[5], N. D. Browning[4], and T. Venkatesan[1]

[1]Center for Superconductivity Research, University of Maryland, College Park, MD 20742-4111, USA.

[2]Department of Materials Science, University of Maryland, College Park, MD 20742-4111, USA.

[3]Department of Computer Science, University of Maryland, College Park, MD 20742-4111, USA.

[4]Lawrence Berkeley National Laboratory, NCEM, One Cyclotron Road, Berkeley, CA 94720, USA

[5]Dept. of Physics, Montana State University, Bozeman, MT 59717, USA

## Abstract

We revisit the most widely investigated and controversial oxide diluted magnetic semiconductor (DMS), Co:TiO$_2$, with a new high temperature film growth, and show that the corresponding material is not only an intrinsic DMS ferromagnet, but also supports a percolative mechanism of ferromagnetism. We establish the uniformity of dopant distribution across the film cross section by Z-contrast imaging via scanning transmission electron microscopy (STEM) and electron energy loss spectroscopy (EELS) at spatial resolution of 0.4 nm and the oxidized 2+ valence state of cobalt by x-ray absorption spectroscopy (XAS). The dependence of magnetic properties on cobalt concentration is consistent with the defect polaron percolation model. The peculiar increase in the transport activation energy above a specific cobalt concentration further emphasizes the polaron contribution to magnetic order.

Spintronics has evoked tremendous interest in recent years due to the prophecy of innovative devices that would be faster, multifunctional, and unique in their own right [1-6]. Vital to the success of these devices is the development of new materials with unique spin functionality. To this end a promising approach being actively explored is to convert a non-magnetic semiconductor into a ferromagnet by introducing a dilute concentration of a magnetic dopant [6-8]. Considerable success has been realized with this approach in the case of Mn doped GaAs [6-12] but the corresponding low Curie temperature (~ 160 K) has raised questions about its potential for widespread use. Therefore the prediction [13] of above-room-temperature ferromagnetism (FM) in Mn:ZnO and the first report [14] of high temperature FM in anatase Co:TiO$_2$ (CTO) have stimulated significant research on oxide diluted magnetic semiconductors (Oxide-DMS) [15-28]. The follow-up research on CTO done by different groups [17-19] questioned the intrinsic origin of FM in this system, especially at high Co concentration (x > 2%) due to the possibility of magnetic cluster formation, either in the form of Co metal and/or Co-Ti-O complex [15]. Recent reports of electric field modulation of the magnetization and coercivity in insulating CTO [24], FM in sputter-grown and UHV-annealed diluted magnetic dielectric CTO [25] and optical magnetic circular dichroism in rutile CTO have stimulated further interest in this material [26]. This most studied oxide-DMS candidate continues to be a controversial subject.

Another interesting issue is the FM exchange mechanism in DMS systems. While it has been shown that itinerant carriers are responsible for mediating the FM exchange [29] in some cases, the precise mechanism in the case of Oxide-DMS is far from clear. It is felt that different and possibly new mechanisms, for example the percolative mechanism suggested by Kaminski and Das Sarma [30], may be operative in such cases. Coey and coworkers [27, 28] have proposed an exchange mechanism that involves a defect state, viz. an F-center.

In this paper we address the two aforementioned issues by exploring the controversial Co:TiO$_2$ system under a new processing condition which has enabled realization of uniform cobalt distribution and ferromagnetism in the anatase TiO$_2$ film at least at low (<~3%) concentration. We have employed scanning transmission electron microscopy (STEM) and electron energy loss spectroscopy (EELS) at 0.14 nm resolution to establish the uniformity of dopant distribution and x-ray absorption spectroscopy (XAS) to establish the oxidized 2+ valence state of cobalt. We also show that the results of Co concentration dependent magnetic and transport properties are in agreement with the computer simulation results of a defect based polaron percolation model.

Thin films of anatase Ti$_{1-x}$Co$_x$O$_{2-d}$ (x = 0 - 0.07) studied in this work were grown on LaAlO$_3$ by pulsed laser deposition at 875 $^o$C in ~ 1-3 x 10$^{-5}$ Torr oxygen. Other deposition parameters were similar to that in Ref. 15. Films having thickness in the range 150 – 800 nm were deposited. The STEM/EELS experiments were carried out using a FEI Tecnai F20 UT microscope operated at 200 kV with a spatial resolution of 0.14 nm in scanning electron transmission mode (0.4 nm for EELS) and an energy resolution of 0.5 eV. High-quality cross section electron microscopy samples were prepared using the shadow technique, a sample preparation technique that combines small angle cleavage technique with focused ion beam (FIB) minimizing sample modification [31]. The XAS measurements were acquired at the MSU X-ray Nanomaterials Characterization Facility located at beamline U4-B of the National Synchrotron Light Source (NSLS), and at beamline 4.0.2 at the Advanced Light Source (ALS). The experimental resolution was 0.4 eV and the spectra were taken at normal incidence and room temperature with linearly polarized soft X-rays in total electron yield mode, measured by sample current with a 250 V extraction grid nearby.

First, we present the evidence for homogeneity of Co. Fig. 1a and 1b show typical transmission electron micrographs for the $Ti_{0.96}Co_{0.04}O_{2-d}$ sample at two different length scales. The micrographs are clearly smooth and featureless, and no clusters of any kind are seen. The distribution of Co in the film was studied by probing the film via tens of EELS line scans across the sample starting with recording one signal on the $LaAlO_3$ substrate and subsequently probing the film in 4 nm steps in direction of the film surface. A typical such line scan is presented in Fig. 1c. The single scans are shifted in the y direction for clarity. The range of energy for the line scan experiments starts at 520 eV and ends at 835 eV in order include the signals of oxygen, cobalt and lanthanum. Proceeding from lower to higher counts in the graph (as indicated by the arrow in Fig.1b), the first signal has been recorded on the $LaAlO_3$ substrate and includes the O K-edge at 532 eV and the La M5-edge at 832 eV. The oxygen signal for the following spectra changes significantly owing to the change in material by crossing the interface and being recorded on the Co doped $TiO_2$ film. In addition, the La M5-edge disappears while the Co signal is present consistently in the film. The white lines for the Co L2,3-edge have been observed at 795 eV and 780 eV respectively implying an oxidized state of Co and the creation of oxygen vacancies in the film. High-resolution Z-contrast imaging throughout the film confirms that no clusters of the dopant are present. Similar characteristics were also seen for all the films studied.

Fig. 2 shows the cobalt L edge XAS spectrum for a 6% cobalt doped anatase sample. This spectrum, as well as those measured for lower concentration samples (not shown), can be compared to those reported by Chambers *et al.* [15] and are clearly indicative of the 2+ oxidation state of cobalt. Thus the high temperature growth condition under a fair amount of oxygen in the ambient offers sufficient oxidizing kinetics to render an oxidized state of cobalt. This result compounded with the dopant uniformity indicates that this system has the characteristics of an intrinsic DMS.

Fig. 3(a-f) show the room temperature magnetic hysteresis data for the films with different cobalt concentrations (x). Note that different scales are used on the figures for clarity. These films clearly have a much lower $M_S$ than that for CTO films grown at lower temperature (having cobalt clusters) as well as the spin only moment of $Co^{2+}$. In Fig. 3g we show the cobalt concentration dependence of the saturation moment ($M_S$) and the coercive field. $M_S$ is seen to decrease systematically with lowering the x. The coercive field shows an interesting non-monotonic dependence on x. The shapes of the hysteresis loops were not found to change much up to the highest temperature of 375 K accessible with SQUID. These facts reinforce absence of any cluster related superparamagnetism in these films, which is not surprising based on our STEM/EELS and XAS results.

Fig. 4a shows the temperature dependent resistivity for the $Ti_{1-x}Co_xO_{2-d}$ films. The observed semiconducting/insulating transport characteristics reflect activated transport. When fitted to activated hopping behavior one obtains the dependence of activation energy on x as shown in Fig. 4b. Fig. 4b also shows the RT resistivity as a function of x. The resistivity is seen to rise slowly with x at low values of x (<1%), but rapidly between 1-2%. It flattens out above 2%. The activation energy is seen to drop slightly at low cobalt concentration (< 0.5%) but shows considerable rise over the range of 1-2%. We will return to the possible cause of this behavior after discussing the relevant model of FM. The observation of FM in these samples with fairly high values of resistivity invites analyses of FM based on the charged-defect polaron formation and magnetic percolation model [27, 30, 32].

Replacement of $Ti^{4+}$ by $Co^{2+}$ causes creation of one oxygen vacancy per Co, leading to two n-type carriers [15]. Note however that for local charge neutrality such oxygen vacancies are primarily localized around $Co^{2+}$. As proposed by Coey and coworkers [27], an electron can be trapped in this vacancy constituting the F-center. The radius of electron orbital is $r = a_0 \varepsilon (m_e/m^*)$, where $a_0$ is Bohr radius, $\varepsilon$ is dielectric constant, $m_e$ is electron mass, and $m^*$ is effective mass in anatase $TiO_2$. For anatase $TiO_2$, the high frequency (static) dielectric constant is ~ 9 (31) and $m^* = m_e$ (ref. 34). For F-center polaron the effective dielectric constant [35, 36] can be obtained as $1/\varepsilon_{eff} = 1/\varepsilon_{hf} - 1/\varepsilon_{static}$. The corresponding value of $\varepsilon_{eff}$ is thus 12.68. Therefore the radius of F-center electron orbit is ~ 6.5 Å, which is about 1.72 lattice units. Since oxygen vacancies reside close to Co, an overlap is expected between the F-center wavefunction and the 3d orbital of Co. Also, depending on 'x', a number of F-centers could overlap with two or more $Co^{2+}$. Now, consider the $Co^{2+} - \otimes - Co^{2+}$ network, where $\otimes$ denotes the F-center. Since $Co^{2+}$ has seven electrons in its outermost orbit, the electron in the overlapping F-center has to have its spin antiparallel to that of the $Co^{2+}$ moment. This will lead to a ferromagnetic coupling between $Co^{2+}$ ions in the $Co^{2+} - \otimes - Co^{2+}$ network. This entity can be envisaged as a polaron for the models of percolative [30, 32] FM in diluted magnetic insulators. In this picture, all the oxygen vacancies may not necessarily form F-centers. Moreover, there may be antiferromagnetic superexchange networks such as $Co^{2+}$ – $O^{2-}$ – $Co^{2+}$. Therefore, we do not expect all nearby Co to be FM-coupled. In fact, as mentioned before, the $M_S$ ($\mu_B$/Co) for all the studied concentrations is lower than $Co^{2+}$ moment. This is similar to the observations by Coey et al.[27] for the case of Fe:$SnO_2$. Also, the number of F-center polarons participating in the percolation network should decrease with decreasing x. Therefore, the $M_S$ can be expected to decrease with the decreasing x, as is indeed observed in our experiments. Note that unlike the transport-percolation-threshold, where a discrete cut off for conduction is observed, a discrete threshold is not expected for $M_S$. Even at very low x there could still be a few finite-sized networks, which may behave as a glassy system giving rise to non-zero $M_S$. Therefore, our observation of magnetic moment for the films with x as low as 0.25 % is not surprising.

We performed computer simulation to capture the essence of these arguments. A cubic lattice of 50 x 50 x 50 sites, with periodic boundary conditions was created and Co atoms were randomly distributed amongst the sites. The Co sites within a distance of the F-center electron wavefunction (1.72 lattice units, based on $\varepsilon_{eff}$ ~ 12.68) were considered to form a bound magnetic polaron complex (BMPC). Following the argument that not all Co ions may have an F-center associated with them (neutral vacancy) or if there are two electrons in the vacancy the local interaction would be antiferromagnetic, some simulations were also performed with only a certain percent fraction of vacancies as effective F-centers for FM. Fig. 5 (main figure, not the inset) shows the distribution of Co in BMPC of different sizes as a function of different cobalt concentrations, for a representative case of F-center fraction of 50%. As the cobalt concentration increases, the fraction of isolated cobalt sites (without FM coupling with any other cobalt) drops rapidly and the fractions of 2, 3, 4…. FM-exchange-coupled cobalt sites rise and decay progressively giving way to larger complexes and finally forming infinite complexes (as in a typical transport percolation problem). In the inset to Fig. 5 we show the cobalt concentration dependence of the fraction of cobalt ions having at least one F-center FM coupled Co ion, for some cases of occupied F-center fractions, along with the experimentally observed saturation moment. If we assume that all cobalt atoms with one or more FM-coupled neighbors will contribute to saturation moment then the data for the case of about 20% effective F-center fraction matches well with the experimental data. In reality one may need a critical size of the BMPC above which a full contribution to the moment could be realized. In this case a higher effective F-center fraction will match with the experimental data.

It is interesting to note that below ~2 % Co, where infinite BMPC are not present (Fig. 5), the films exhibit a coercive field $H_C$ (see Fig. 3g) that decreases with decreasing x. This can be attributed to the increase in the fraction of superparamagnetic BMPCs (zero coercive field) with decreasing x. Small $H_C$ is also seen for the magnetically well connected system at high dopant concentration having infinite BMPCs, since the magnetization reversal could be easier in such case. With these competing behaviors in the low and high concentration regimes, the $H_C$ is expected to peak in the intermediate concentration regime, as observed (Fig. 3g). The changes in the hysteresis loop shape as a function of x [Figs. 3(a – f)] which also correspond to the specific nature of the coercive field dependence on magnetization thus appear to relate well to the connectivity issue ascribable to the development of percolation network. Indeed, the observed curve has a striking similarity with that for the well-studied problem of FM in a system wherein a FM phase is embedded in a non-magnetic medium [37, 38]. In such systems, the coercive field peaks near the percolation threshold, suggesting the possible validity of percolation scenario in the present case. The occurrence of a jump in resistivity and the activation energy in the neighborhood of 1-2% Co, where maximum changes occur in the percolation networks as shown by the simulations, also supports the polaron picture [39]. Indeed the peculiar behavior of activation energy further suggests a magnetic contribution to the polaron transport. If it were only for lattice distortions, the overlap of polaronic wavefunctions with increasing cobalt content (and associated vacancies for charge neutrality) is expected to soften the barrier for hopping, which may be the cause of the initial small drop when magnetic interactions are yet to develop. On the other hand if a hop involves disturbance of spin order, once it is set up, a cost of the order of exchange energy J may have to be involved [39]. Some key local states contributing to such hop processes are shown as inset to Fig. 4b. These considerations which are peculiar to the F-center model and relate to the corresponding occupancy also emphasize the sensitivity of oxide-DMS FM to growth conditions.

In conclusion, we have presented micro-structural, chemical and magnetic evidence that under the new high temperature growth conditions, Co distributes uniformly in the anatase $TiO_2$ matrix at low cobalt concentrations. Our data and computer simulation suggest that F-center based polaron formation and their percolation are responsible for FM.

The authors acknowledge support under DARPA grant # N000140210962 and NSF-MRSEC DMR-00-80008. The authors also acknowledge fruitful discussions with S. Das Sarma.

References:

*e-mail: shinde@squid.umd.edu

¶ e-mail: ogale@squid.umd.edu

# Currently at the Department of Physics, University of California, Berkeley.

Figure captions:

Figure 1: (a, b) Typical transmission electron micrographs for the $Ti_{0.96}Co_{0.04}O_{2-d}$ sample at different length scales, (c) An EELS line scan across the cross section of the sample: data recorded in steps of 4 nm.

Figure 2: The cobalt L edge x-ray absorption (XAS) spectrum for anatase $Ti_{0.94}Co_{0.06}O_{2-d}$ film.

Figure 3: (a-f) Room temperature magnetic hysteresis data for the films with different cobalt concentrations (x%), (g) The cobalt concentration dependence of the saturation moment ($M_S$) and the coercive field.

Figure 4: (a) The temperature dependent resistivity for the $Ti_{1-x}Co_xO_{2-d}$ films, (b) The dependence of activation energy and $\rho_{300K}$ on x. Inset shows some local states involved in magnetic contribution to hopping transport.

Figure 5: Results of computer simulation study based on the F-center polaron model for a representative case of 50% effective F-centers : The curves show the cobalt concentration dependence of the fraction of isolated cobalt ions, as well as cobalt ions with polaronic overlap with 2 or 3, 4 or more, and infinite number of cobalt ions. Inset shows the cobalt concentration dependence of the fraction of cobalt ions that have one or more F-center polaron coupled neighbor(s), for three cases of effective F-center percent fraction (full curve 100%, dashed curve 50%, dotted curve 20%). The experimental data on saturation moment are also plotted.

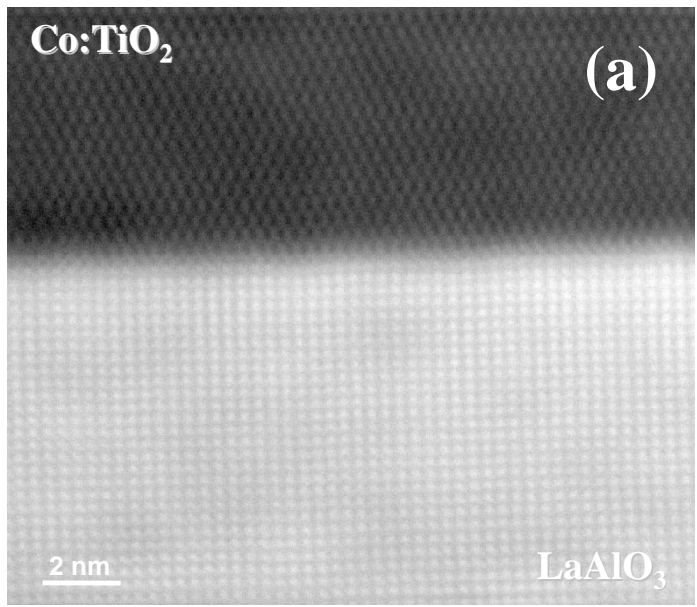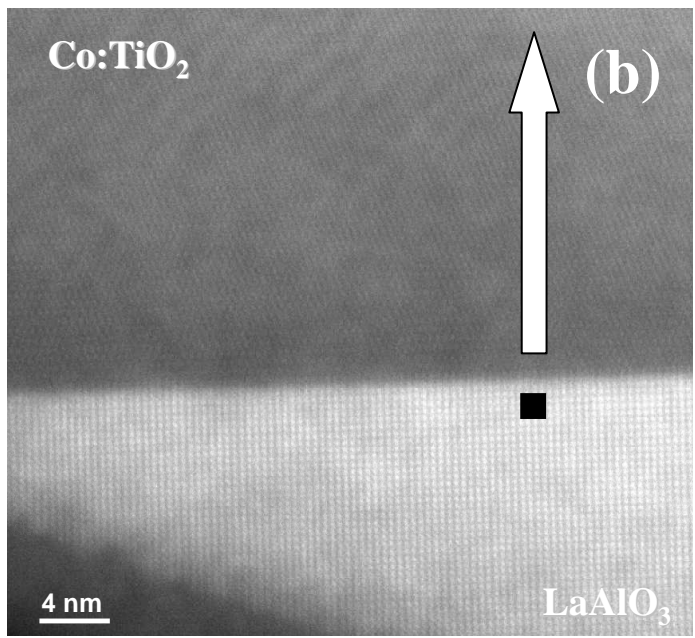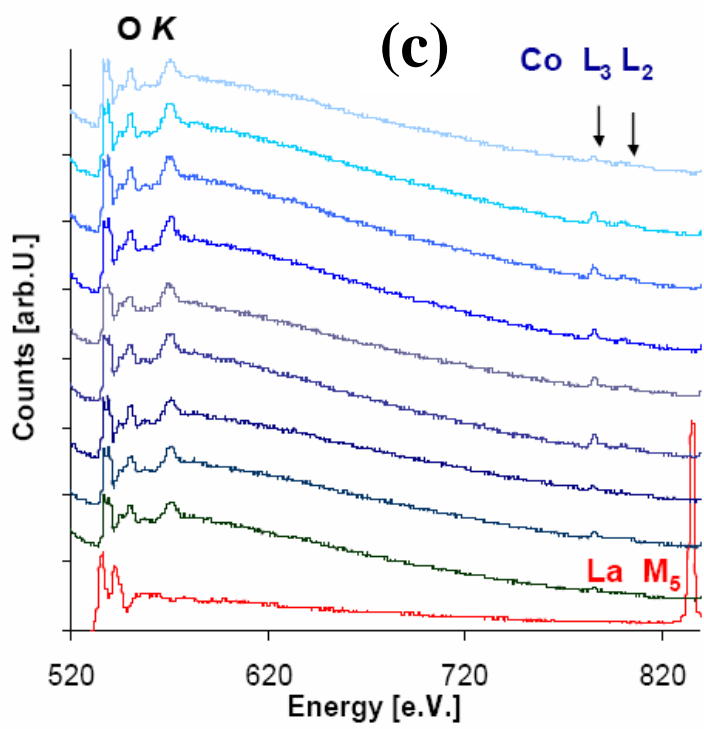

Fig. 1

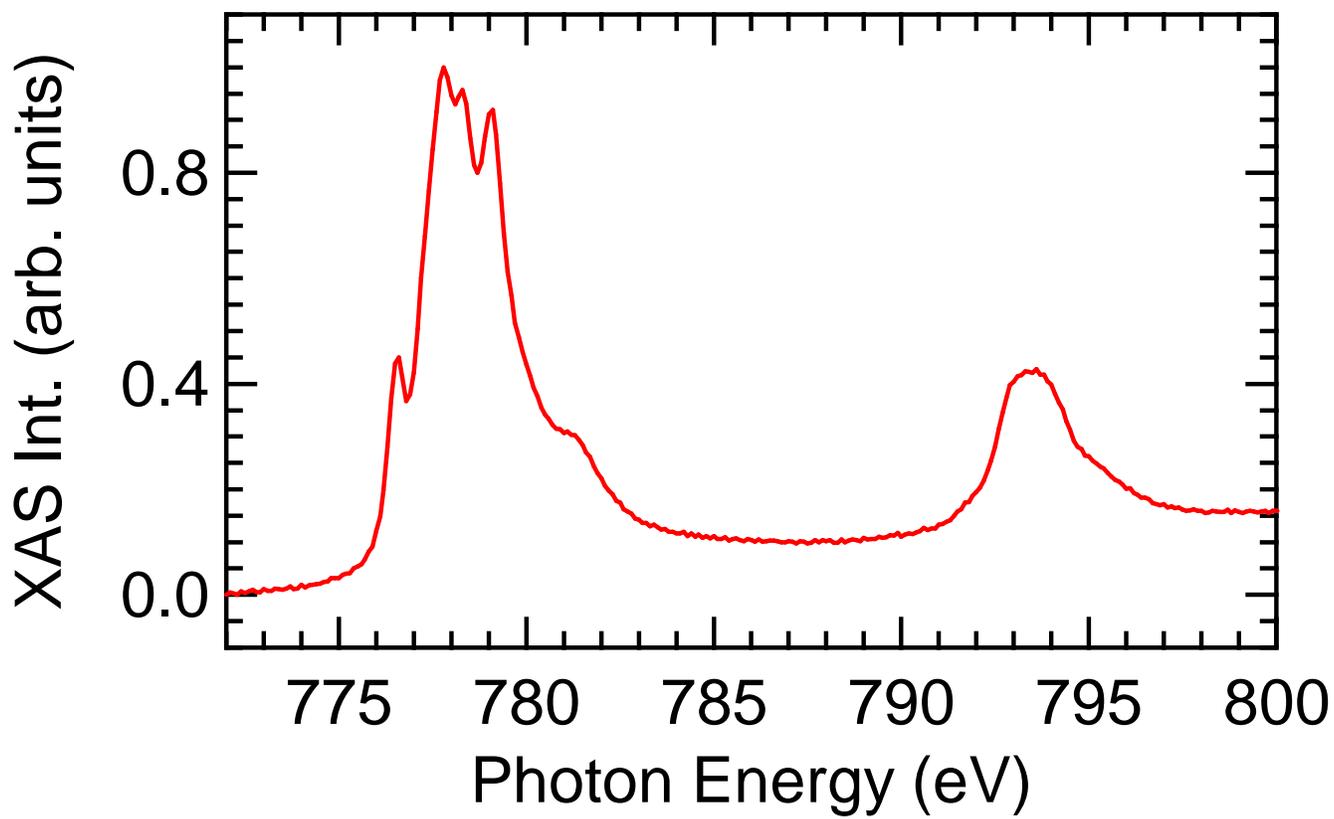

Fig. 2

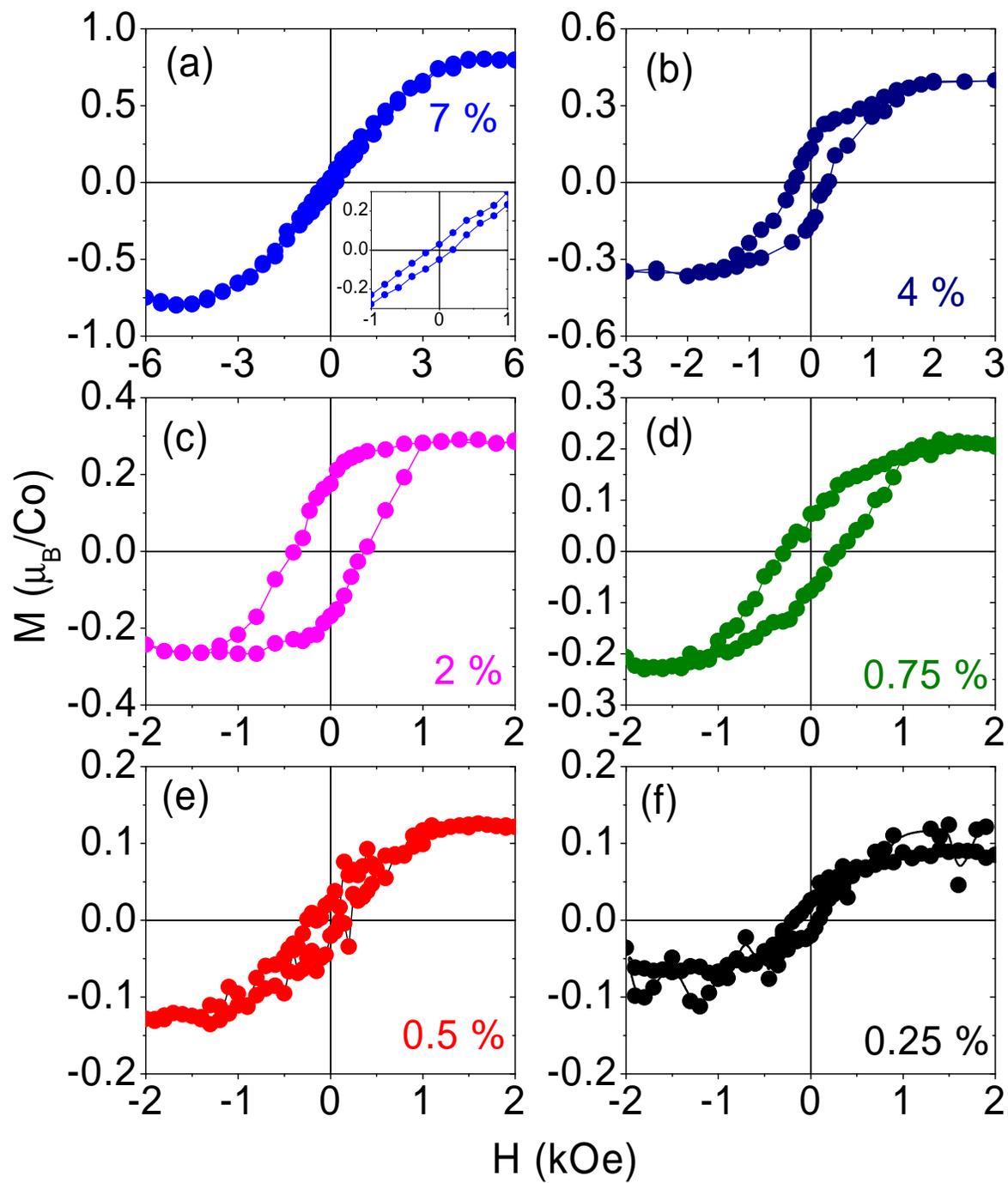

Fig. 3 a-f

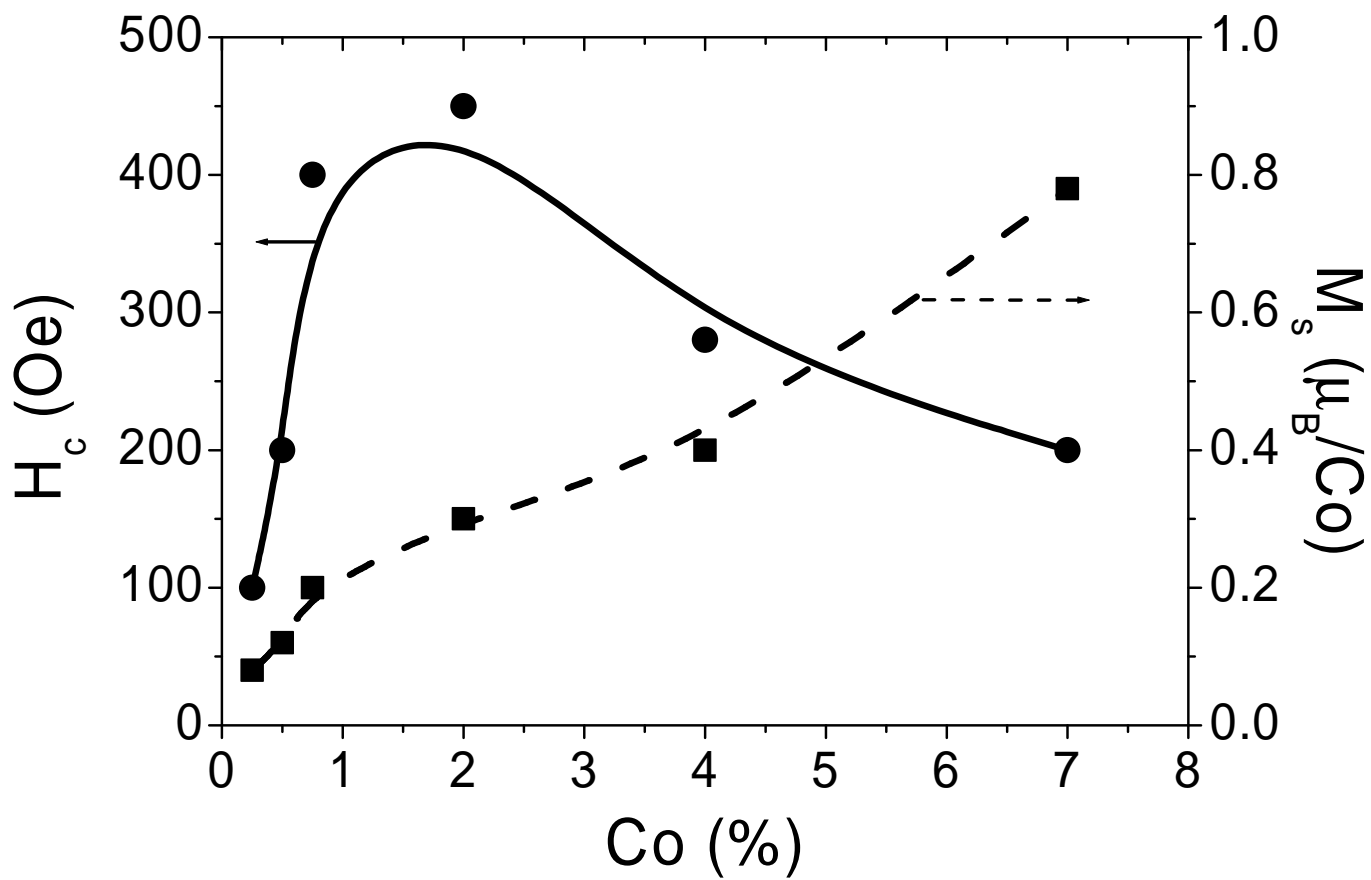

Fig. 3 g

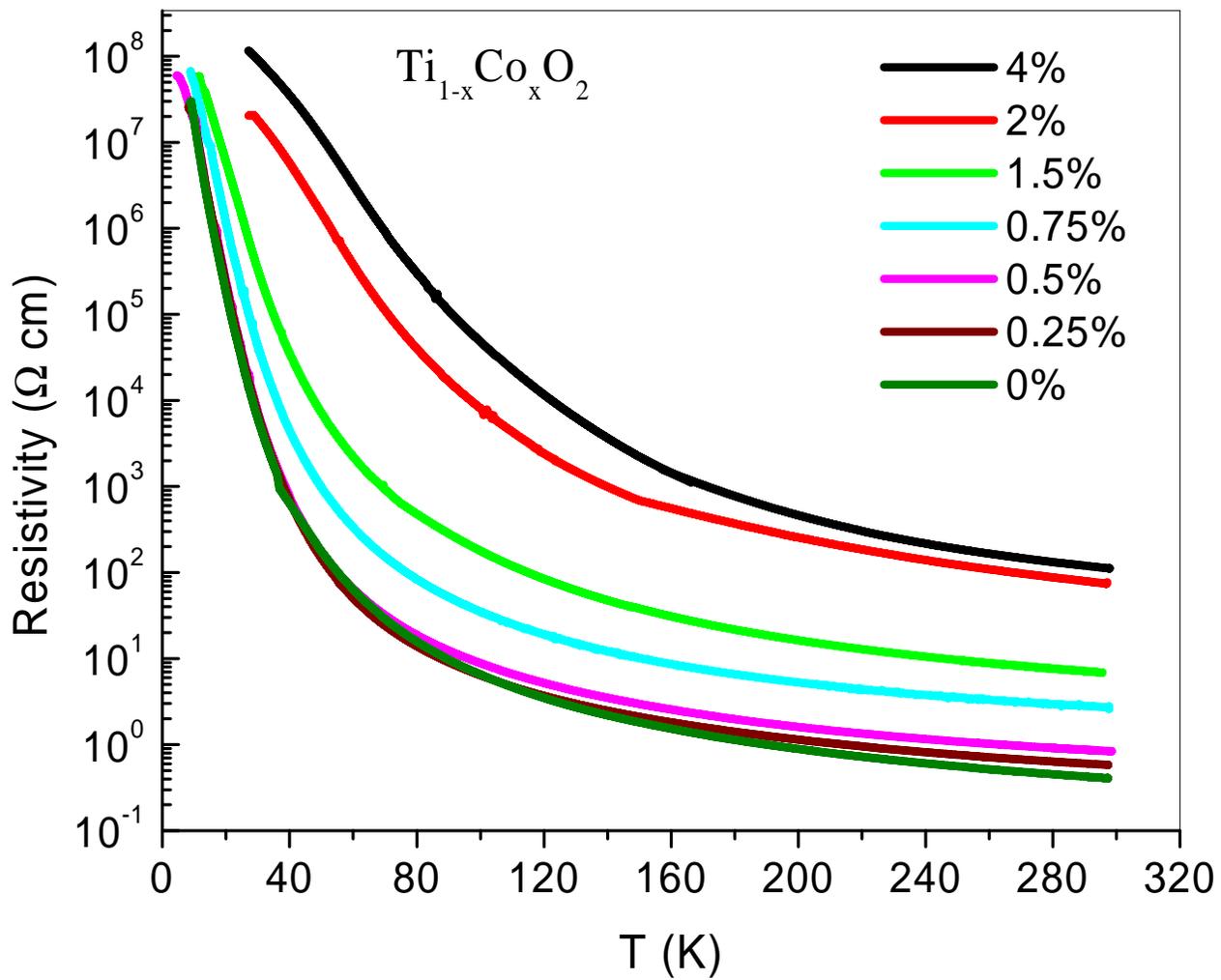

Fig. 4 a

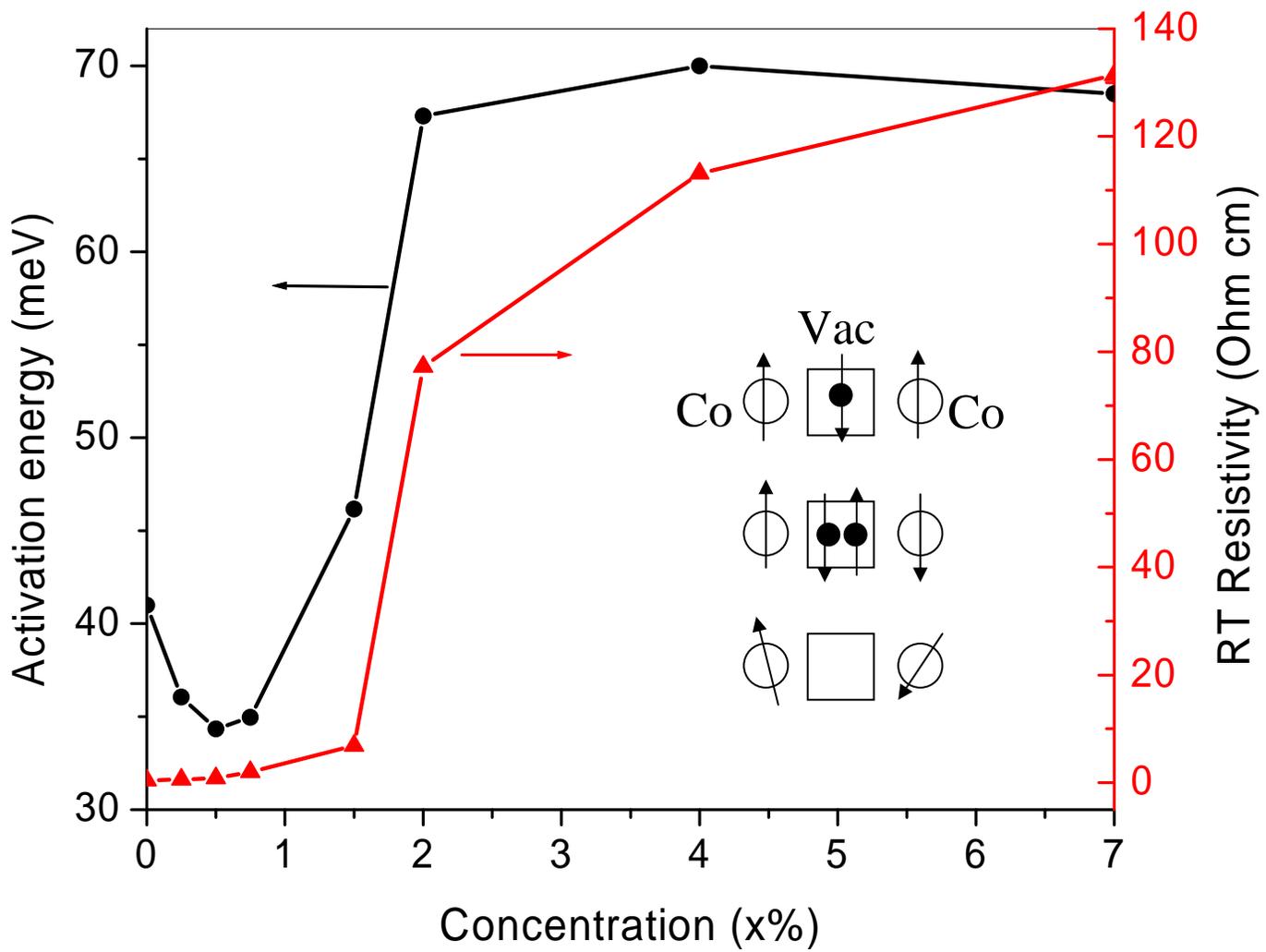

Fig. 4 b

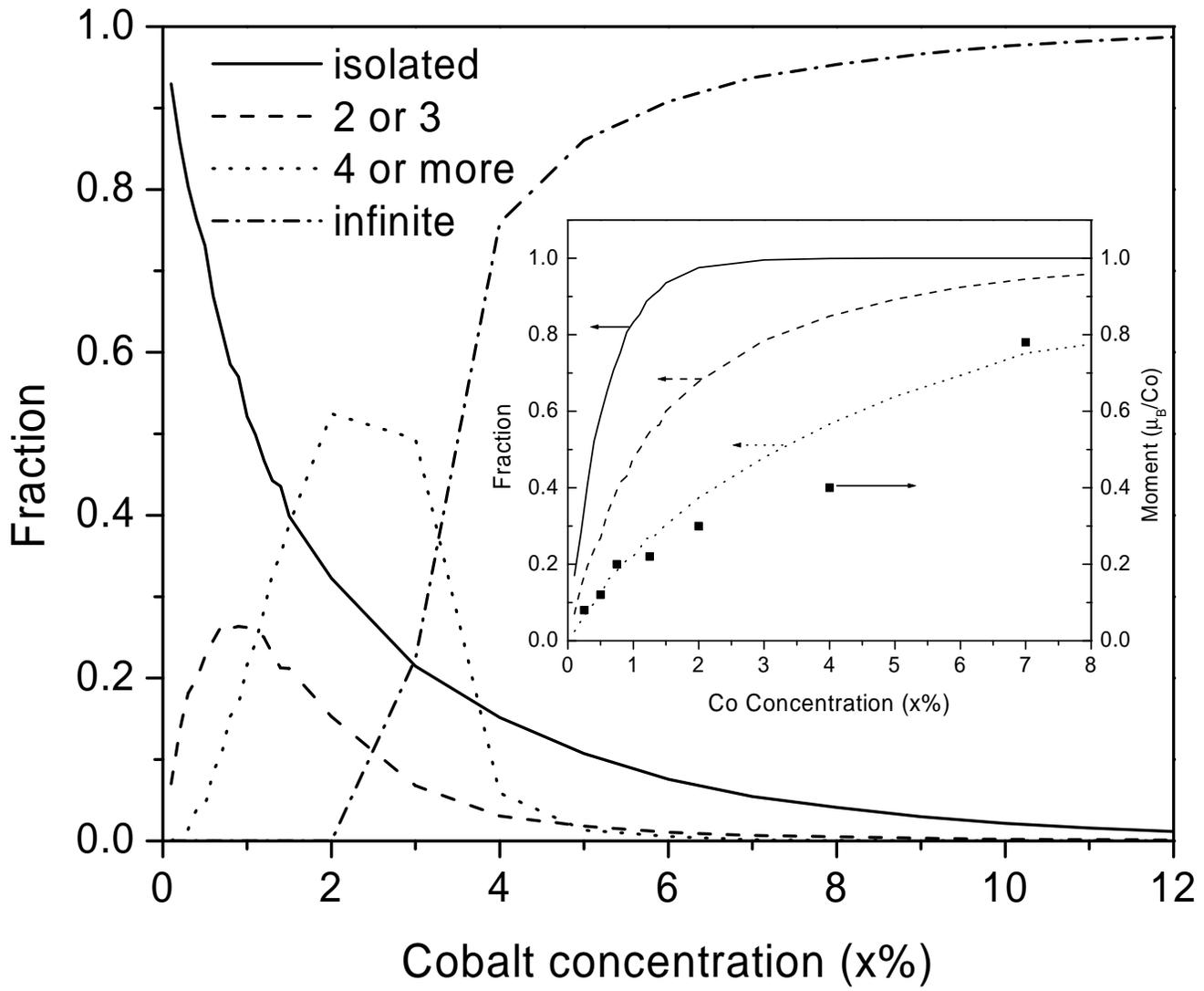

Fig. 5